\documentclass[aps,pra,twocolumn,showpacs]{revtex4}
\usepackage{bbm,amsmath,amssymb,amsbsy,graphicx,subfigure,times}

\usepackage[latin1]{inputenc}
\newcommand{\R}{\mathbbm{R}}

\newcommand{\be}{\begin{equation}}
\newcommand{\ee}{\end{equation}}
\newcommand{\bea}{\begin{eqnarray}}
\newcommand{\eea}{\end{eqnarray}}

\newcommand{\sy}[1]{Sp_{(#1,\R)}}

\newcommand{\tr}{{\rm Tr}\,}
\renewcommand{\det}{{\rm Det}\,}
\newcommand{\gr}[1]{\boldsymbol{#1}}
\newcommand{\ket}[1]{|#1\rangle}
\newcommand{\bra}[1]{\langle#1|}

\newcommand{\N}{{\cal N}}

\newcommand{\sig}{\gr{\sigma}}

\newcommand{\bet}{\gr{\beta}}
\newcommand{\alp}{\gr{\alpha}}

\newcommand{\eq}[1]{Eq.~(\ref{#1})}
\newcommand{\ineq}[1]{Ineq.~(\ref{#1})}

\newcommand{\ie}{\emph{i.e.}~}

\begin{document}
\title{Gaussian measures of entanglement versus negativities: \\
the ordering of two--mode Gaussian states}
\date{June 20, 2005}
\author{Gerardo Adesso and Fabrizio Illuminati}
\affiliation{Dipartimento di Fisica ``E. R. Caianiello'',
Universit\`a di Salerno; CNR-Coherentia, Gruppo di Salerno; \\ and
INFN Sezione di Napoli-Gruppo Collegato di Salerno, Via S. Allende,
84081 Baronissi (SA), Italy}


\begin{abstract}
In this work we study the entanglement of general (pure or mixed)
two--mode Gaussian states of continuous variable systems by
comparing the two available classes of {\it computable} measures
of entanglement: entropy-inspired Gaussian convex-roof measures,
and PPT-inspired measures (negativity and logarithmic negativity).
We first review the formalism of Gaussian measures of
entanglement, adopting the framework introduced in [M.~M.~Wolf
{\em et al.}, Phys. Rev. A {\bf 69}, 052320 (2004)], where the
Gaussian entanglement of formation was defined. We compute
explicitely Gaussian measures of entanglement for two important
families of nonsymmetric two--mode Gaussian states, namely the
states of extremal (maximal and minimal) negativities at fixed
global and local purities, introduced in [G.~Adesso {\em et al.},
Phys. Rev. Lett. {\bf 92}, 087901 (2004)]. This analysis allows to
compare the different {\em orderings} induced on the set of
entangled two--mode Gaussian states by the negativities and by the
Gaussian measures of entanglement. We find that in a certain range
of values of the global and local purities (characterizing the
covariance matrix of the corresponding extremal states), states of
minimum negativity can have more Gaussian entanglement of
formation than states of maximum negativity. Consequently,
Gaussian measures and negativities are definitely inequivalent
measures of entanglement on nonsymmetric two--mode Gaussian
states, even when restricted to a class of extremal states. On the
other hand, the two families of entanglement measures are
completely equivalent on symmetric states, for which the Gaussian
entanglement of formation coincides with the true entanglement of
formation. Finally, we show that the inequivalence between the two
families of continuous-variable entanglement measures is somehow
limited. Namely, we rigorously prove that, at fixed negativities,
the Gaussian measures of entanglement are bounded from below.
Moreover, we provide some strong evidence suggesting that they are
as well bounded from above.
\end{abstract}
\maketitle

\section{Introduction}

Quantum information with continuous variables (CV)
\cite{book,brareview} is a flourishing field dedicated to the
manipulation of the information using quantum states
governed by the laws of quantum mechanics. This approach contrasts
with the usual methods involving discrete-spectrum observables
(such as, e. g., polarization, spin, energy level) of
single photons, atoms or ions. The ability of quantum
states with continuous spectra to implement
quantum cryptography \cite{cvcrypto}, quantum teleportation
\cite{cvtelep,furuscience,dernier,bra00,telepoppate}, entanglement
swapping \cite{cvswap,dernier}, dense coding \cite{cvdense}, quantum
state storage \cite{memory}, and, to some extent, quantum computation
\cite{cvqc} processes, brings up new and exciting perspectives.

The crucial resource enabling a better-than-classical manipulation
and processing of information is CV entanglement, introduced for the
first time in the landmark paper by Einstein, Podolski and Rosen
\cite{epr} in 1935. There, it was shown that the simultaneous
eigenstate of relative position and total momentum of two particles
(or of a two modes of the radiation field) contains perfect quantum
correlations, {\ie}infinite CV entanglement. While this state is
clearly an unphysical, unnormalizable state, it can be approximated
arbitrarily well by two-mode squeezed Gaussian states with large
enough squeezing parameter. The special class of Gaussian states
(which includes thermal, coherent, and squeezed states), thus
emerges quite naturally in the CV scenario. These entangled states
can be easily produced and manipulated experimentally, and moreover
their mathematical description is greatly simplified due to the fact
that, while still living in a infinite-dimensional Hilbert space,
their relevant properties (such as entanglement and mixedness) are
completely determined by the finite-dimensional covariance matrix of
two-point correlations between the canonically conjugated quadrature
operators. Therefore, clarifying the characterization and
quantification of CV entanglement in two-mode and, eventually,
multimode Gaussian states stands as a major issue in the field of CV
quantum information, as the amount of entanglement contained in a
certain state directly quantifies its usefulness for information and
communication tasks like teleportation \cite{telepoppate}.

For the prototypical entangled states of a CV system,
the two--mode Gaussian states, much is known about entanglement
qualification, as the separability is completely
characterized by the necessary and sufficient PPT criterion
(positivity of the partially transposed state) \cite{simon00}),
and also with regard to its quantifcation. Concerning
the latter aspect, the negativity (quantifying the violation of
the necessary and sufficient PPT condition for separability)
is {\it computable} for all two--mode Gaussian states \cite{logneg}.
Moreover, for symmetric two-mode Gaussian states also the entanglement
of formation is computable \cite{giedke03}, and it turns out to be
completely equivalent to the negativity for these states.

Another measure of CV entanglement, adapted for the class of
Gaussian states, has been introduced in Ref. \cite{GEOF}, where
the Gaussian entanglement of formation (an upper bound to the true
entanglement of formation) was defined as the cost of producing an
entangled mixed state out of an ensemble of pure, Gaussian states.
While the Gaussian entanglement of formation coincides with the true
one for symmetric states, at present it is not known whether this equality
holds for nonsymmetric states as well \cite{openprob}.

In this work, aimed at sheding new light on the quantification of
entanglement in two--mode Gaussian states, we compute the
Gaussian entanglement of formation and, in general, the family of
Gaussian entanglement measures, for two different classes of two--mode
Gaussian states, namely the states of extremal, maximal and minimal,
negativities at fixed global and local purities \cite{prl,extremal}.
We find that the two families of entanglement measures
(negativities and Gaussian measures) are not equivalent for
nonsymmetric states. Remarkably, they may induce a completely
different {\em ordering} on the set of entangled two--mode Gaussian
state: a nonsymmetric state $\varrho_A$ can be more entangled than
another state $\varrho_B$, with respect to negativities, and less
entangled than the same state $\varrho_B$, with respect to Gaussian
measures of entanglement. However, the inequivalence between the two families of
measures is somehow bounded: we show that, at fixed
negativities, the Gaussian entanglement measures are rigorously
bounded from below. Moreover, we provide strong evidence hinting
that they should be bounded from above as well.

The paper is organized as follows. In Sec. \ref{secgauss} we set up
the notation and review the basic properties of Gaussian states of CV
systems. In Sec. \ref{secent} we review the main results on the
characterization of separability in Gaussian states, introducing
also two families of measures of entanglement, respectively the
negativities and the Gaussian entanglement measures. In Sec.
\ref{secgem} we compute the latter for two--mode Gaussian states,
solving the problem explicitely for the states of extremal
negativities at fixed purities, described in Sec. \ref{secextr}. In
Sec. \ref{secorder} we compare the orderings induced by negativities
and Gaussian measures on the set of extremal two--mode Gaussian
states. In Sec. \ref{secvs} we compare the two families of measures
for generic two--mode Gaussian states, finding lower and upper
bounds on one of them, when keeping the other fixed. Finally, in
Sec. \ref{secconcl} we summarize our results and discuss future
perspectives.

\section{Gaussian states: Definitions and notation}\label{secgauss}
A continuous variable (CV) system is described by a Hilbert space
${\cal H}=\bigotimes_{i=1}^{n} {\cal H}_{i}$ resulting from the
tensor product structure of infinite dimensional Fock spaces ${\cal H}_{i}$'s.
Let $a_{i}$ be the annihilation operator acting on ${\cal H}_{i}$,
and $\hat q_{i}=(a_{i}+a^{\dag}_{i})$ and $\hat
p_{i}=(a_{i}-a^{\dag}_{i})/i$ be the related quadrature phase
operators. The corresponding phase space variables will be denoted
by $q_{i}$ and $p_{i}$. Let $\hat X = (\hat q_{1},\hat
p_{1},\ldots,\hat q_{n},\hat p_{n})$ denote the vector of the
operators $\hat q_{i}$ and $\hat p_{i}$. The canonical commutation
relations for the $\hat X_{i}$ can be expressed in terms of the
symplectic form ${\Omega}$
\[
[\hat X_{i},\hat X_j]=2i\Omega_{ij} \; ,
\]
\[
{\rm with}\quad{\Omega}\equiv \bigoplus_{i=1}^{n} {\omega}\; , \quad
{\omega}\equiv \left( \begin{array}{cc}
0&1\\
-1&0
\end{array}\right) \; .
\]

The states of a CV system can be equivalently described by a positive
trace-class operator (the density matrix $\varrho$) or by
quasi--probability distributions such as the Wigner function
\cite{barnett}. States with Gaussian characteristic functions and
quasi--probability distributions are referred to as Gaussian states.
Such states are at the heart of information processing in CV systems
\cite{brareview} and are the subject of our analysis. By definition,
a Gaussian state $\varrho$ is completely characterized by the first
and second statistical moments of the quadrature field operators, which will be
denoted, respectively, by the vector of first moments $\bar
X\equiv\left(\langle\hat X_{1} \rangle,\langle\hat
X_{1}\rangle,\ldots,\langle\hat X_{n}\rangle, \langle\hat
X_{n}\rangle\right)$ and the covariance matrix (CM) $\sig$ of
elements
\begin{equation}
\sig_{ij}\equiv\frac{1}{2}\langle \hat{X}_i \hat{X}_j + \hat{X}_j
\hat{X}_i \rangle - \langle \hat{X}_i \rangle \langle \hat{X}_j
\rangle \, , \label{covariance}
\end{equation}
where, for any observable $\hat{o}$, the expectation value
$\langle\hat o\rangle\equiv\,{\rm Tr}(\varrho\hat o)$. Notice that
the entries of the CM can be expressed as energies by multiplying
them by the quantity $\hbar \omega$, where $\omega$ is the frequency
of the considered mode. In fact, for any $n$-mode state (even non
Gaussian) the quantity $\hbar \omega\tr{\sig}/4$ is simply the
average of the non interacting Hamiltonian $\sum_{i=1}^{n}
(a^{\dag}_ia_i + 1/2)$. First moments can be arbitrarily adjusted by
local unitary operations (displacements), which cannot affect any
property related to entropy or entanglement. Therefore, they will be
unimportant to the present scope and we will set them to $0$ in the
following, without any loss of generality.

The canonical commutation relations and the positivity of the
density matrix $\varrho$ imply
\begin{equation}
\sig+ i\Omega\ge 0 \; , \label{bonfide}
\end{equation}
Inequality (\ref{bonfide}) is the necessary and sufficient
constraint the matrix $\sig$ has to fulfill to be a CM corresponding
to a physical Gaussian state \cite{simon87,simon}. More in general,
the previous condition is necessary for the CM of {\em any},
generally non Gaussian, state. We note that such a constraint
implies $\sig\ge0$.

A major role in the theoretical and experimental manipulation of
Gaussian states is played by unitary operations which preserve the
Gaussian character of the states on which they act. Such operations
are all those generated by Hamiltonian terms at most quadratic in
the field operators.  As a consequence of the Stone-Von Neumann
theorem, any such unitary operation at the Hilbert space level
corresponds, in phase space, to a symplectic transformation, {\ie}to
a linear transformation $S$ which preserves the symplectic form
$\Omega$, so that $\Omega=S^T \Omega S$. Symplectic transformations
on a $2n$-dimensional phase space form the (real) symplectic group
$\sy{2n}$. Such transformations act linearly on first moments and by
congruences on covariance matrices: $\sig\mapsto S^{\sf T} \sig S$.
One has $\det{S}=1$, $\forall\,S\in\sy{2n}$. Ideal beam splitters,
phase shifters and squeezers are all described by some kind of
symplectic transformation. A particularly important symplectic
transformation is the one realizing the decomposition of a Gaussian
state in normal modes. Through this decomposition, thanks to
Williamson theorem \cite{williamson36}, the CM of a $n$--mode
Gaussian state can always be written in the so-called Williamson
normal, or diagonal form
\begin{equation}
\sig=S^T \gr\nu S \; , \label{willia}
\end{equation}
where $S\in Sp_{(2n,\mathbb{R})}$ and $\gr\nu$ is the CM
\begin{equation}
\gr\nu=\,{\rm diag}({\nu}_{1},{\nu}_{1},\ldots,{\nu}_{n},{\nu}_{n})
\, , \label{therma}
\end{equation}
corresponding to a tensor product of thermal states with a diagonal
density matrix $\varrho^{_\otimes}$ given by
\be
\varrho^{_\otimes}=\bigotimes_{i}
\frac{2}{\nu_{i}+1}\sum_{k=0}^{\infty}\left(
\frac{\nu_{i}-1}{\nu_{i}+1}\right)\ket{k}_{i}{}_{i}\bra{k}\; ,
\label{thermas} \ee
where $\ket{k}_i$ denotes the number state of order $k$
in the Fock space ${\cal H}_{i}$.

The quantities $\nu_{i}$'s form the symplectic spectrum of the CM
$\sig$, and they can be computed as the eigenvalues of the matrix
$|i\Omega\sig|$. Such eigenvalues are in fact invariant under the
action of symplectic transformations on the matrix $\sig$. The
symplectic eigenvalues $\nu_{i}$ encode essential informations on
the Gaussian state $\sig$ and provide powerful, simple ways to
express its fundamental properties. For instance, in terms of the
symplectic eigenvalues $\nu_{i}$, the uncertainty relation
(\ref{bonfide}) reads \be {\nu}_{i}\ge1 \; . \label{sympheis}
\ee

Moreover, the entropic quantities of Gaussian states can be as well
expressed in terms of their symplectic eigenvalues and invariants
\cite{extremal}. Notably, the purity $\tr{\varrho^2}$ of a Gaussian
state $\varrho$ is simply given by the symplectic invariant
$\det{\sig}=\prod_{i=1}^{n}\nu_i^2$, being \cite{mga}
\begin{equation}\label{purity}
\mu \equiv \tr{\varrho^2} = \frac{1}{\sqrt{\det{\sig}}}\,.
\end{equation}

\subsection{Two--mode states}
This work is focused on two--mode Gaussian states: we thus briefly
review here some of their basic properties. The expression of the
two--mode CM $\sig$ in terms of the three $2\times 2$ matrices
$\alp$, $\bet$, $\gr\gamma$, that will be useful in the following,
takes the form
\begin{equation}
\sig\equiv\left(\begin{array}{cc}
{\alp}&{\gr\gamma}\\
{\gr\gamma}^{\sf T}&{\bet}
\end{array}\right)\, . \label{espre}
\end{equation}
For any two--mode CM ${\sig}$ there is a local symplectic operation
$S_{l}=S_{1}\oplus S_{2}$ which brings ${\sig}$ in the so called
standard form ${\sig}_{sf}$ \cite{simon00, duan00}
\begin{equation}
S_{l}^{T}{\sig}S_{l}={\sig}_{sf} \equiv \left(\begin{array}{cccc}
a&0&c_{+}&0\\
0&a&0&c_{-}\\
c_{+}&0&b&0\\
0&c_{-}&0&b
\end{array}\right)\; . \label{stform}
\end{equation}
States whose standard form fulfills $a=b$ are said to be symmetric.
Let us recall that any pure state ($\mu = 1$) is symmetric and
fulfills $c_{+}=-c_{-}=\sqrt{a^2-1}$. The correlations $a$, $b$,
$c_{+}$, and $c_{-}$ are determined by the four local symplectic
invariants ${\rm Det}{\sig}=(ab-c_{+}^2)(ab-c_{-}^2)$, ${\rm
Det}{\alp}=a^2$, ${\rm Det}{\bet}=b^2$, ${\rm
Det}{\gr\gamma}=c_{+}c_{-}$. Therefore, the standard form
corresponding to any CM is unique (up to a common sign flip in $c_-$
and $c_{+}$).

For two--mode states, the uncertainty principle
Ineq.~(\ref{bonfide}) can be recast as a constraint on the
$Sp_{(4,{\mathbb R})}$ invariants ${\rm Det}\sig$ and
$\Delta(\sig)={\rm Det}{\alp}+\,{\rm Det}{\bet}+2 \,{\rm
Det}{\gr\gamma}$ \cite{serafozzi}:
\begin{equation}
\Delta(\sig)\le1+\,{\rm Det}\sig \label{sepcomp}\; .
\end{equation}

The symplectic eigenvalues of a two--mode Gaussian state will be
denoted as $\nu_{-}$ and $\nu_{+}$, with $\nu_{-}\le \nu_{+}$, with the
 uncertainty relation (\ref{sympheis}) reducing to \be \label{symptwo} \nu_{-}\ge 1 \; .
\ee A simple expression for the $\nu_{\mp}$ can be found in terms of
the two $Sp_{(4,\mathbb{R})}$ invariants (invariants under global,
two--mode symplectic operations) \cite{logneg, serafozzi}
\begin{equation}
2{\nu}_{\mp}^2=\Delta(\sig)\mp\sqrt{\Delta^2(\sig) -4\,{\rm
Det}\,\sig} \, . \label{sympeig}
\end{equation}

\section{Entanglement of Gaussian states}\label{secent}

In this section we recall the main results on the qualification and
quantification of entanglement for Gaussian states of CV systems.

\subsection{Qualification: PPT criterion}

The positivity of the partially transposed state (Peres-Horodecki
PPT criterion \cite{PPT}) is necessary and sufficient for the
separability of two--mode Gaussian states \cite{simon00} and, more
generally, of all $(1+n)$--mode Gaussian states under $1\times
n$-mode bipartitions \cite{werwolf} and of symmetric and
bisymmetric $(m+n)$--mode Gaussian states under $m\times n$-mode bipartitions
\cite{unitarily}. In general, the partial transposition
$\tilde{\varrho}$ of a bipartite quantum state $\varrho$ is defined
as the result of the transposition performed on only one of the two
subsystems in some given basis. In phase space, the action of
partial transposition amounts to a mirror reflection of one of the
four canonical variables \cite{simon00}. The CM $\sig$ is then
transformed into a new matrix $\tilde\sig$ which differs from $\sig$
by a sign flip in ${\rm Det}\,\gr\gamma$. Therefore the invariant
$\Delta(\sig)$ is changed into $\tilde{\Delta}({\sig})
\equiv\Delta(\tilde{\sig})=\,{\rm Det}\,{\alp}+ \,{\rm
Det}\,{\bet}-2\,{\rm Det}\,{\gr\gamma}$. Now, the symplectic
eigenvalues $\tilde{\nu}_{\mp}$ of $\tilde{\sig}$ read \be
\tilde{\nu}_{\mp}=
\sqrt{\frac{\tilde{\Delta}(\sig)\mp\sqrt{\tilde{\Delta}^2(\sig)
-4\,{\rm Det}\,\sig}}{2}} \, . \label{sympareig} \ee The PPT
criterion for separability thus reduces to a simple inequality that
must be satisfied by the smallest symplectic eigenvalue
$\tilde{\nu}_{-}$ of the partially transposed state \be
\tilde{\nu}_{-}\ge 1 \: , \label{symppt} \ee which is equivalent to
\be \tilde{\Delta}(\sig)\le \,{\rm Det}\,\sig+1 \; . \label{ppt} \ee
Moreover, the above inequalities imply ${\rm Det}\,{\gr\gamma}=c_{+}c_{-}<0$
as a necessary condition for a two--mode Gaussian state to be
entangled. Therefore, the quantity $\tilde{\nu}_{-}$ encodes all the
qualitative characterization of the entanglement for arbitrary (pure
or mixed) two--mode Gaussian states.

\subsection{Negativities} \label{secnega}

From a quantitative point of view, a measure of entanglement which
can be computed for general Gaussian states is provided by the {\em
negativity} $\N$, first introduced in Ref.~\cite{zircone}, later
thoroughly discussed and extended in Refs.~\cite{logneg,jenstesi} to
CV systems. The negativity of a quantum state $\varrho$ is defined
as \be {\cal N}(\varrho)=\frac{\|\tilde \varrho \|_1-1}{2}\: , \ee
where $\tilde\varrho$ is the partially transposed density matrix and
$\|\hat o\|_1=\,{\rm Tr}|\hat o|$ stands for the trace norm of the
hermitian operator $\hat o$. The quantity ${\cal N} (\varrho)$ is
equal to $|\sum_{i}\lambda_{i}|$, the modulus of the sum of the
negative eigenvalues of $\tilde\varrho$, quantifying the extent to
which $\tilde\varrho$ fails to be positive. Strictly related to $\N$
is the {\em logarithmic negativity} $E_{\N}$, defined as
$E_{\N}\equiv \log \|\tilde{\varrho}\|_{1}$, which constitutes an
upper bound to the {\em distillable entanglement} of the quantum
state $\varrho$ and is related to the entanglement cost under PPT
preserving operations \cite{auden03}. Both the negativity and the
logarithmic negativity have been proven to be monotone under LOCC
(local operations and classical communications)
\cite{logneg,jenstesi,plenio05}, a crucial property for a {\em bona
fide} measure of entanglement. Moreover, the logarithmic negativity
possesses the nice property of being additive.

For any two--mode Gaussian state $\varrho$ it is easy to show that
both the negativity and the logarithmic negativity are
simple decreasing functions of $\tilde{\nu}_{-}$
\cite{logneg,extremal} \be
\|\tilde\varrho\|_{1}=\frac{1}{\tilde{\nu}_{-}}\;\Rightarrow
\N(\varrho)=\max \, \left[
0,\frac{1-\tilde{\nu}_{-}}{2\tilde{\nu}_{-}} \right] \, , \ee \be
\label{ennu} E_{\N}(\varrho)=\max\,\left[0,-\log
\tilde{\nu}_{-}\right] \, . \ee These expressions directly quantify
the amount by which the necessary and sufficient PPT condition
(\ref{symppt}) for separability is violated. The symplectic
eigenvalue $\tilde{\nu}_{-}$ thus completely qualifies and
quantifies (in terms of negativities) the entanglement of a
two--mode Gaussian state $\sig$: for $\tilde \nu_- \ge 1$ the state
is separable, otherwise it is entangled. Finally, in
the limit of vanishing $\tilde \nu_-$, the
negativities grow unboundedly.

\subsection{Entanglement of Formation}

In the special instance of symmetric two--mode Gaussian states,
the {\em entanglement of formation} (EoF) \cite{bennet96}, can be
computed as well \cite{giedke03}. We recall that the EoF $E_{F}$ of a
quantum state $\varrho$ is defined as \be
E_{F}(\varrho)=\min_{\{p_i,\ket{\psi_i}\}}\sum_i p_i E(\ket{\psi_i})
\; , \label{eof} \ee where the minimum is taken over all the pure
states realizations of $\varrho$:
\[
\varrho=\sum_i p_i \ket{\psi_i}\bra{\psi_i} \; .
\]
The asymptotic regularization of the
entanglement of formation coincides with the {\em entanglement cost}
$E_C (\varrho)$, defined as the minimum number of singlets
(maximally entangled antisymmetric two-qubit states) which is needed
to prepare the state $\varrho$ through LOCC \cite{ecost}.

The optimal convex decomposition of \eq{eof} has been found for
symmetric two--mode Gaussian states, and turns out to be Gaussian,
that is, the absolute minimum is realized within the set of pure two--mode
Gaussian states \cite{giedke03},
yielding \be E_F = \max\left[ 0,h(\tilde{\nu}_{-}) \right] \; ,
\label{eofgau} \ee with
\begin{equation}\label{hentro}
h(x)=\frac{(1+x)^2}{4x}\log \left[\frac{(1+x)^2}{4x}\right]-
\frac{(1-x)^2}{4x}\log \left[\frac{(1-x)^2}{4x}\right].
\end{equation}
Such a quantity is, again, a monotonically decreasing function of
$\tilde{\nu}_{-}$, thus providing a quantification of the
entanglement of symmetric states {\em equivalent} to the one
provided by the negativities.

As a consequence of this equivalence, it is tempting to
conjecture that there exists a unique quantification
of entanglement for two--mode Gaussian states, embodied by the smallest
symplectic eigenvalue $\tilde \nu_-$ of the partially transposed CM,
and that the different measures simply provide trivial
rescalings of the same unique quantification.
In particular, the {\em ordering} induced on the set of
entangled Gaussian state is uniquely defined for the subset
of symmetric two--mode states, and it is independent of the
chosen measure of entanglement. However, regrettably,
in Sec. \ref{secorder} we will indeed show that
different measures of entanglement induce, in general,
different orderings on the set of nonsymmetric
two--mode Gaussian states.

\subsection{Gaussian convex-roof extended measures}

In this subsection we consider a family of entanglement measures
exclusively defined for Gaussian states of CV systems. The formalism
of {\em Gaussian entanglement measures} (Gaussian EMs) has been
introduced in Ref. \cite{GEOF} where the Gaussian EoF has been
defined and analyzed. Furthermore, the framework developed in Ref.
\cite{GEOF} is general and enables to define generic
Gaussian EMs of bipartite entanglement by applying the Gaussian
convex roof, that is, the convex roof over pure Gaussian decompositions
only, to any {\em bona fide} measure of bipartite entanglement defined
for pure Gaussian states. The original motivation for the introduction
of Gaussian EMs stems from the unfortunate fact that the
optimization problem \eq{eof} for the computation of the EoF of
nonsymmetric two--mode Gaussian states has not yet been solved, and
it stands as an open problem in the theory of entanglement
\cite{openprob}. However, the task can be somehow simplified by restricting
to decompositions into pure Gaussian states only. The resulting
measure, named as Gaussian EoF in Ref. \cite{GEOF}, is an upper bound
to the true EoF and coincides with it for symmetric two--mode Gaussian
states.

In general, we can define a Gaussian EM $G_E$ as follows. For any
pure Gaussian state $\psi$ with CM $\sig^P$, one has
\begin{equation}\label{Gaussian EMp}
G_E (\sig^P) \equiv E(\psi)\,,
\end{equation}
where $E$ can be {\em any} proper measure of entanglement
of pure states, defined as a monotonically increasing function of the
entropy of entanglement ({\ie}the von Neumann entropy of the reduced
density matrix of one party).

For any mixed Gaussian state $\varrho$ with CM $\sig$, one has
\cite{GEOF}
\begin{equation}\label{Gaussian EMm}
G_E (\sig) \equiv \inf_{\sig^P \le \sig} G_E(\sig^P)\,.
\end{equation}
If the function $E$ is taken to be exactly the entropy of
entanglement, then the corresponding Gaussian EM is known as {\em
Gaussian EoF} \cite{GEOF}. In Ref. \cite{jensata} the properties of
the Gaussian EoF have been further investigated, and interesting
connections with the capacity of bosonic Gaussian channels have been
established.

In general, the definition \eq{Gaussian EMm} involves an
optimization over all pure Gaussian states with CM $\sig^P$ smaller
than the CM $\sig$ of the mixed state whose entanglement one wishes
to compute. Despite being a simpler optimization problem than that
appearing in the definition \eq{eof} of the true EoF  (which, in
CV systems, would imply considering decompositions over all,
Gaussian {\it and} non-Gaussian pure states), the Gaussian EMs
cannot be expressed in a simple closed form, not even in the
simplest instance of (nonsymmetric) two--mode Gaussian states. It is
the aim of the present paper to compute Gaussian EMs for two
relevant classes of, {\it generally nonsymmetric},
two--mode Gaussian states, namely the states of
{\em extremal} (maximal and minimal) negativity at fixed global and
local purities \cite{prl,extremal}, which will be reviewed in Sec.
\ref{secextr}. This will provide an insight into the problem of the
ordering \cite{ordering} of two--mode Gaussian states with respect
to different measures of entanglement, leading to results somehow
similar to those obtained for systems of two qubits \cite{frank},
where in general the EoF and the negativity are found to be inequivalent.

Before moving on to the explicit computations, let us recall, as an important
side remark, that any Gaussian EM is an entanglement monotone under
Gaussian LOCC. The proof given in Sec. IV of Ref. \cite{GEOF} for
the Gaussian EoF, in fact, automatically extends to every Gaussian EM
constructed via the Gaussian convex roof of any proper measure $E$ of
pure-state entanglement.

\section{Gaussian entanglement measures \newline for two--mode Gaussian
states} \label{secgem}

The problem of evaluating Gaussian EMs for a generic two--mode
Gaussian state has been solved in Ref. \cite{GEOF}. However, the
explicit result contains so ``cumbersome'' expressions (involving
the solutions of a fourth-order algebraic equation), that the authors
of Ref. \cite{GEOF} considered them not particularly useful to be
reported explicitely in their paper.

We recall here the computation procedure \cite{GEOF} that we will
need in the following. For any two-mode Gaussian state with CM $\sig
\equiv \sig_{sf}$ in standard form \eq{stform}, a generic Gaussian
EM $G_E$ is given by the entanglement $E$ of the least entangled
pure state with CM $\sig^P \le \sig$. Denoting by $\gamma_q$
(respectively $\gamma_p$) the $2 \times 2$ submatrix obtained from
$\sig$ by canceling the even (resp. odd) rows and columns, we have,
explicitely
\begin{equation}\label{cpcq}
\gamma_q = \left(
\begin{array}{ll}
  a & c_+ \\
  c_+ & b \\
\end{array}
\right)\,,\quad \gamma_p = \left(
\begin{array}{ll}
  a & c_- \\
  c_- & b \\
\end{array}
\right)\,.
\end{equation}
All the covariances relative to the ``position'' operators of the
two modes are grouped in $\gamma_q$, and analogously for the
``momentum'' operators in $\gamma_p$. The total CM can then be
written as a direct sum $\sig = \gamma_q \oplus \gamma_p$.
Similarly, the CM of a generic pure two--mode Gaussian state in
standard form (it has been proven that the CM of the optimal pure
state has to be in standard form as well \cite{GEOF}) can be written
as $\sig^P = \gamma_q^P \oplus \gamma_p^P$, where the global purity
of the state imposes $(\gamma_p^P)^{-1} = \gamma_q^P \equiv \Gamma$.
The pure states involved in the definition of the Gaussian EM must
thus fulfill the condition
\begin{equation}\label{rim}
\gamma_p^{-1} \le \Gamma \le \gamma_q\,.
\end{equation}

This problem is endowed with a nice geometric description
\cite{GEOF}. Writing the matrix $\Gamma$ in the basis constituted by
the identity matrix and the three Pauli matrices,
\begin{equation}\label{Gamma} \Gamma = \left(
\begin{array}{cc}
  x_0 + x_3 & x_1 \\
  x_1 & x_0 - x_3 \\
\end{array}
\right)\,,
\end{equation}
the expansion coefficients $(x_0,x_1,x_3)$ play the role of
space-time coordinates in a three-dimensional Minkowski space. In
this picture, for example, the rightmost inequality in \eq{rim} is
satisfied by matrices $\Gamma$ lying on a cone, which is equivalent
to the (backwards) light cone of $C_q$ in the Minkowski space; and
similarly for the leftmost inequality. Indeed, one can show that,
for the optimal pure state $\sig^P_{opt}$ realizing the minimum in
\eq{Gaussian EMm}, the two inequalities in \eq{rim} have to be
simultaneously saturated \cite{GEOF}. From a geometrical point of
view, the optimal $\Gamma$ has then to be found on the rim of the
intersection of the forward and the backward cones of
$\gamma_p^{-1}$ and $\gamma_q$, respectively. This is an ellipse,
and one is left with the task of minimizing the entanglement $E$ of
$\sig^P = \Gamma \oplus \Gamma^{-1}$ (see \eq{Gaussian EMp}) for
$\Gamma$ lying on this ellipse \cite{noteole}.

At this point, let us pause to briefly recall that any pure two--mode Gaussian
state $\sig^P$ is locally equivalent to a two--mode squeezed state
with squeezing parameter $r$, described by a CM
\begin{equation}\label{cm2ms}
\sig^P_{sq} = \left(\begin{array}{cccc}
\cosh (2r)&0& \sinh (2r)&0\\
0&\cosh (2r)&0&-\sinh (2r)\\
\sinh(2r)&0&\cosh(2r)&0\\
0&-\sinh(2r)&0&\cosh(2r)
\end{array}\right)\,.
\end{equation}
The following statements are then equivalent: (i) $E$ is a
monotonically increasing function of the entropy of entanglement;
(ii) $E$ is a monotonically increasing function of the single--mode
determinant $m\equiv\det\gr\alpha \equiv \det\gr\beta$ (see
\eq{espre}); (iii) $E$ is a monotonically decreasing function of the
local purity $\mu_i\equiv\mu_1\equiv\mu_2$ (see \eq{purity}); (iv)
$E$ is a monotonically decreasing function of the smallest
symplectic eigenvalue $\tilde\nu_-^P$ of the partially transposed CM
$\tilde\sig^P$; (v) $E$ is a monotonically increasing function of
the squeezing parameter $r$. This chain of equivalences is immediately
proven by simply recalling that a pure state is completely specified by
its single--mode marginals, and that for a single--mode Gaussian state
there is a unique symplectic invariant (the determinant), so that all
conceivable entropic quantities are monotonically increasing
functions of this invariant \cite{extremal}. In particular, statement
(ii) allows us to minimize
directly the single--mode determinant over the ellipse:
\begin{equation}\label{mdef}
m = 1 + \frac{x_1}{\det \Gamma}\,,
\end{equation}
with $\Gamma$ given by \eq{Gamma}.

To simplify the
calculations, one can move to the plane of the ellipse with a
Lorentz boost which preserves the relations between all the cones;
one can then choose the transformation so that the ellipse degenerates
into a circle (with fixed radius), and introduce polar coordinates
on this circle.
The calculation of the Gaussian EM for any two--mode
Gaussian state is thus finally reduced to the minimization of $m$ from
\eq{mdef}, at given standard-form covariances of $\sig$, as a
function of the polar angle $\theta$ on the circle \cite{noteole}.
So far, this technique has been applied to the computation of
the Gaussian EoF by minimizing \eq{mdef} {\em numerically} \cite{GEOF} (see also
\cite{oleposter}). In addition to that, as already mentioned, the
Gaussian EoF has been exactly computed for symmetric states, and it
has been proven that in this case the Gaussian EoF is the true EoF
\cite{giedke03}.

In this work we present new analytical calculations of
the Gaussian EMs for two relevant classes of nonsymmetric two--mode
Gaussian states: the states of {\em extremal} negativities at fixed
global and local purities \cite{extremal}, which will be introduced
in the next subsection.
We begin by writing the general expression of
the single--mode determinant \eq{mdef} in terms of the covariances of a
generic two--mode state (see \eq{stform}) and of the polar angle
$\theta$. After some tedious but straightforward
algebra, one finds
\begin{widetext}
\begin{eqnarray}\label{mfunc}
m_\theta (a,b,c_+, c_-)\ =\ 1 &+&
\left\{\left[c_+(ab-c_-^2)-c_-+\cos \theta \sqrt{\left[a -
b(ab-c_-^2)\right]\left[b-a(ab-c_-^2)\right]}\right]^2\right\}
\nonumber \\
& \times & \Bigg\{
2\left(ab-c_-^2\right)\left(a^2+b^2+2c_+c_+ \right)  \nonumber \\
& &\ -\ \frac{\cos \theta\left[2abc_-^3+\left(a^2+
b^2\right)c_+c_-^2+\left(\left(1-2b^2\right)a^2+
b^2\right)c_--ab\left(a^2+b^2- 2\right)c_+\right]}{\sqrt{\left[a -
b(ab-c_-^2)\right]\left[b-a(ab-c_-^2)\right]}} \nonumber \\
& &\ +\ \sin \theta\left(a^2-
b^2\right)\sqrt{1-\frac{\left[c_+(ab-c_-^2)+c_-\right]^2}{\left[a -
b(ab-c_-^2)\right]\left[b-a(ab-c_-^2)\right]}} \, \Bigg\}^{-1}\,,
\end{eqnarray}
\end{widetext}
where we have assumed $c_+ \ge |c_-|$ without any loss of
generality. This implies that, for any entangled state, $c_+ > 0$
and $c_- < 0$. The Gaussian EM (defined in terms of the function $E$
on pure states, see \eq{Gaussian EMp}) of a generic two--mode
Gaussian state coincides then with the entanglement
$E$ computed on the pure state with
$m=m_{opt}$, with $m_{opt} \equiv \min_\theta (m_\theta)$.
Accordingly, the
symplectic eigenvalue $\tilde \nu_-$ of the partial transpose of the
corresponding optimal pure-state CM $\sig^{P}_{opt}$, realizing the
infimum in \eq{Gaussian EMm}, would read (see Eq. \eq{sympareig})
\begin{equation}\label{nutopt}
\tilde\nu_{- opt}^{P} \equiv \tilde \nu_-(\sig^{P}_{opt}) =
\sqrt{m_{opt}} - \sqrt{m_{opt}-1}\,.
\end{equation}
As an example, for the Gaussian EoF one has \be \label{geofm}
G_{E_F}(\sig) = h\left(\tilde\nu_{- opt}^{P}(m_{opt})\right)\,,\ee
with $h(x)$ defined by \eq{hentro}.

Finding the minimum of \eq{mfunc} analytically for a generic state
is a difficult task. Numerical investigations show that the equation
$\partial_\theta m_\theta = 0$ can have from one to four physical
solutions (in a period) corresponding to extremal points, and the
global minimum can be attained in any of them depending on the
parameters of the CM $\sig$ under inspection. However, a closed
solution can be found for two important classes of nonsymmetric
two--mode Gaussian states, as we will now show.

\subsection{Parametrization of two--mode covariance matrices and
definition of extremal
states}   \label{secextr}

We have shown in Refs. \cite{prl,extremal} that, at fixed global
purity $\mu \equiv \tr{\varrho^2}$ of the global state $\varrho$,
and at fixed local purities $\mu_{1,2} \equiv \tr{\varrho_{1,2}^2}$
of each of the two reduced single--mode states $\varrho_i = {\rm Tr}_{j
\neq i} \varrho$, the smallest symplectic eigenvalue $\tilde \nu_-$ of the partial
transpose of the CM $\sig$ of a generic two--mode Gaussian state
(which qualifies its separability by the PPT criterion, and
quantifies its entanglement in terms of the negativities) is strictly
bounded from above and from below. This entails the existence of
two disjoint classes of extremal states, namely the states of maximum
negativity for fixed global and local purities (GMEMS), and the states
of minimum negativity for fixed global and local purities (GLEMS)
\cite{extremal}. The negativities of the two {\em extremal} classes
of Gaussian states, moreover, turn out to remain very close to each other
for all the possible assignments of the three purities, allowing for a
reliable experimental estimate of the negativity of a generic two-mode
Gaussian state in terms of the {\em average negativity} \cite{prl}.
The latter is determined by knowledge of the three purities alone,
which, in turn, may be experimentally measured in direct, possibly
efficient, ways \cite{fiuracerf}.

Recalling these results, one can provide a very useful and insightful
parametrization of the {\em entangled} two--mode Gaussian states in
standard form (see also \cite{polacchi}). In fact, the coefficients
appearing in \eq{stform} can be rewritten, in general, according to
the following, useful parametrization:
\begin{widetext}
\begin{eqnarray}
  a &=& s + d\,,\qquad b\,\,=\,\,s-d\,,\label{asd} \\
  c_{\pm} &=& \frac{1}{4\sqrt{s^2 -
            d^2}} \left\{\sqrt{\left[4 d^2 + \frac{1}{2} \left(g^2 +
                          1\right) (\lambda - 1) - \left(2 d^2 +
                          g\right) (\lambda + 1)\right]^2 -
            4 g^2} \right. \nonumber \\
            & \pm & \left.\sqrt{\left[4 s^2 + \frac{1}{2} \left(g^2 +
                          1\right) (\lambda - 1) - \left(2 d^2 +
                          g\right) (\lambda + 1)\right]^2 - 4
                          g^2}\right\} \label{cpm}\,,
\end{eqnarray}
\end{widetext}
where the two local purities are regulated by the parameters $s$
and $d$, being $\mu_1 = (s+d)^{-1},\,\mu_2 = (s-d)^{-1}$, and the
global purity is $\mu = g^{-1}$. The coefficient $\lambda$ embodies
the only remaining degree of freedom needed for the complete determination
of the negativities, once the three purities have been fixed.
It ranges from the minimum $\lambda = -1$ (corresponding to the GLEMS) to
the maximum $\lambda = +1$ (corresponding to the GMEMS).
Therefore, as it varies, $\lambda$ encompasses all possible entangled
two--mode Gaussian states compatible with a given set of assigned values
of the purities. The constraints that the parameters $s,\,d,\,g$ must obey
for \eq{stform} to denote a proper CM of a physical state are: $s \ge 1$,
$|d| \le s-1$, and
\begin{equation}\label{gmbound}
g \ge 2 |d| + 1\,,
\end{equation}

If the global purity is large enough so that \ineq{gmbound} is
saturated, GMEMS and GLEMS coincide, the CM becomes independent of
$\lambda$, and the two classes of extremal states coalesce
into a unique class, completely determined by the marginals $s$ and $d$.
We denote these states as GMEMMS \cite{extremal}, that is,
Gaussian two--mode states of maximal negativity at fixed local purities.
Their CM is simply characterized by $c_{\pm} = \pm \sqrt{s^2 -(d+1)^2}$, where we have
assumed, without any loss of generality, that $d \ge 0$ (corresponding to
choose, for instance, mode $1$ as the more mixed one: $\mu_1 \le \mu_2$).

In general \cite{prl}, a GMEMS ($\lambda = +1$) is entangled for
\be\label{gmement} g < 2s-1\,, \ee while a GLEMS ($\lambda = -1$) is
entangled for smaller $g$, namely \be\label{glement} g < \sqrt{2(s^2 +
d^2) -1}\,. \ee To have a physical insight on these peculiar
two--mode states, let us recall \cite{extremal} that GMEMS are simply
nonsymmetric thermal squeezed states, usually referred to as
maximally entangled mixed states in CV systems. On the other hand,
GLEMS are mixed states of partial minimum uncertainty, in the sense
that the smallest symplectic eigenvalue of their CM is equal to $1$,
saturating the uncertainty inequality (\ref{symptwo}).

We are now equipped with the necessary tools, and in the next
subsection we move on to compute Gaussian EMs for the two extremal
classes of nonsymmetric two--mode Gaussian states, the GLEMS and the GMEMS.

\subsection{Gaussian entanglement of minimum-negativity states (GLEMS)}

We want to find the optimal pure state $\sig^P_{opt}$ entering in
the definition \eq{Gaussian EMm} of the Gaussian EM. To do this, we
have to minimize the single--mode determinant of $\sig^P_{opt}$,
given by \eq{mfunc}, over the angle $\theta$. It turns out that, for
a generic GLEMS, the coefficient of $\sin \theta$ in the last line
of \eq{mfunc} vanishes, and the expression of the single--mode
determinant reduces to the simplified form
\begin{equation}\label{mglems}
m_\theta^{_{\rm  GLEMS}} = 1 + \frac{[A \cos \theta + B]^2}{2(a b -
c_-^2)[(g^2 - 1) \cos \theta  + g^2 + 1]}\,,
\end{equation}
with $A = c_+(a b - c_-^2) + c_-,\, B = c_+(a b - c_-^2) - c_-,$ and
$a,b,c_\pm$ are the covariances of GLEMS, obtained from
Eqs.~(\ref{asd},\ref{cpm}) setting $\lambda = -1$.

The only relevant solutions (excluding the unphysical and the
trivial ones) of the equation $\partial_\theta m_\theta = 0$ are
$\theta = \pi$ and $$\theta = \pm \theta^\ast \equiv \arccos
\left[\frac{3+g^2}{1-g^2} - \frac{2c_-}{c_+(a b - c_-^2) + c_-}
\right]\,.$$ Studying the second derivative $\partial^2_\theta
m_\theta$ for $\theta = \pi$ one finds immediately that, for
\begin{equation}\label{gminmax} g \ge
\sqrt{-\frac{2c_+(a b - c_-^2) + c_-}{c_-}}
\end{equation}
(remember that $c_- \le 0$), the solution $\theta = \pi$ is a
minimum. In this range of parameters, the other solution $\theta =
\theta^\ast$ is unphysical (in fact $|\cos \theta^\ast| \ge 1$), so
$m_{\theta=\pi}$ is the global minimum. When, instead,
\ineq{gminmax} is violated, $m_\theta$ has a local maximum for
$\theta = \pi$ and two minima appear at $\theta = \pm \theta^\ast$.
The global minimum is attained in any of the two, given that,
for GLEMS, $m_\theta$ is invariant under reflection with respect to the axis
$\theta = \pi$. Collecting, substituting, and simplifying the
obtained expressions, we arrive at the final result for the optimal
$m$:
\begin{widetext}
\begin{equation}\label{m2glems}
m_{opt}^{_{\rm GLEMS}}=\left\{
\begin{array}{ll}
    1, & g \ge \sqrt{2(s^2 + d^2)-1} \qquad {\hbox{[separable state]}}\; ; \\
& \\
    \frac{16 s^2 d^2}{(g^2-1)^2}, & \sqrt{\frac{\left(4 s^2 + 1\right) d^2 + s^2 +
            4 s \sqrt{\left(s^2 + 1\right) d^2 + s^2} |d|}{d^2 + s^2}}
\le g < \sqrt{2(s^2 + d^2)-1}\; ; \\
& \\
    \frac{-g^4 +
      2 \left(2 d^2 + 2 s^2 +
            1\right) g^2 - \left(4 d^2 - 1\right) \left(4 s^2 -
            1\right) - \sqrt{\delta}}{8 g^2}, & 2 |d| + 1 \le g < \sqrt{\frac{\left(4 s^2 + 1\right) d^2 + s^2 +
            4 s |d| \sqrt{\left(s^2 + 1\right) d^2 + s^2} }{d^2 + s^2}}\,. \\
\end{array}
\right.
\end{equation}
\end{widetext}
Here $\delta \equiv (2 d - g - 1) (2 d - g + 1) (2 d + g - 1) (2 d +
                g + 1) (g - 2 s - 1) (g - 2 s + 1) (g + 2 s - 1) (g + 2 s +
                1)$.

Immediate inspection crucially reveals that $m_{opt}^{_{\rm GLEMS}}$ is {\em
not} in general a function of the symplectic eigenvalue
$\tilde\nu_-$ alone. Therefore, unfortunately, the Gaussian EMs, and
in particular, the Gaussian EoF, are not equivalent to the
negativities for GLEMS. Further remarks will be given in the
following, when the Gaussian EMs of GLEMS and GMEMS will be compared
and their relationship with the negativities will be elucidated.

\subsection{Gaussian entanglement of maximum-negativity states (GMEMS)}

The minimization of $m_\theta$ from \eq{mfunc} can be carried out in
a  simpler way in the case of GMEMS, whose covariances can be retrieved
from \eq{cpm} setting $\lambda = 1$. First of all, one
can notice that, when expressed as a function of the Minkowski
coordinates $(x_0,x_1,x_3)$, corresponding to the submatrix $\Gamma$
\eq{Gamma} of the pure state $\sig^P = \Gamma \oplus \Gamma^{-1}$
entering in the optimization problem \eq{Gaussian EMm}, the
single--mode determimant $m$ of $\sig^P$ is globally minimized for
$x_3=0$. In fact, from \eq{mdef}, $m$ is minimal, with respect to
$x_3$, when $\det\Gamma = x_0^2 - x_1^2 - x_3^2$ is maximal. Next,
one can show that for GMEMS there always exists a matrix $\Gamma$,
with $x_3=0$, which is a simultaneous solution of the two matrix
equations obtained by imposing the saturation of the two sides of
inequality (\ref{rim}). As a consequence of the above discussion,
this matrix would denote the optimal pure state $\sig^P_{opt}$.
Solving the system of equations $\det(\gamma_q - \Gamma) = \det(\Gamma - \gamma_p^{-1}) =
0$, where the matrices involved are explicitely defined combining
\eq{cpcq} and \eq{cpm} with $\lambda = 1$, one finds the following
two solutions for the coordinates $x_0$ and $x_1$:
\begin{equation}
\begin{split}
\!\!x_0^{\pm} &= \frac{(g + 1) s \pm \sqrt{\left[(g - 1)^2 - 4
d^2\right] \left(-d^2 + s^2 - g\right)}}{2 \left(d^2 + g\right)}\,,\\
\!\!x_1^{\pm} &= \frac{(g + 1) \sqrt{-d^2 + s^2 - g}\pm s \sqrt{(g -
1)^2 - 4 d^2}}{2 \left(d^2 + g\right)}\,.
\end{split}
\end{equation}
The corresponding pure state $\sig^{P\pm} = \Gamma^{\pm} \oplus
{\Gamma^{\pm}}^{-1}$ turns out to be, in both cases, a two--mode
squeezed state described by a CM of the form \eq{cm2ms}, with $\cosh
(2r) = x_0^\pm$. Because the single--mode determinant $m = \cosh^2
(2r)$ for these states, the optimal $m$ for GMEMS is simply equal to
$(x_0^-)^2$. Summarizing,
\begin{equation}\label{m2gmems}
m_{opt}^{_{\rm GMEMS}}=\left\{
\begin{array}{ll}
    1, \qquad \qquad g \ge 2s-1 \quad {\hbox{[separable state]}}\; ; \\
& \\
    \frac{\left\{(g + 1) s - \sqrt{\left[(g - 1)^2 - 4 d^2\right] \left(-d^2 +
                      s^2 - g\right)}\right\}^2}{4 \left(d^2 + g\right)^2},\\
\qquad \qquad \ \ \; 2 |d| + 1 \le g  < 2s-1\; . \\
\end{array}
\right.
\end{equation}
Once again, also for the class of GMEMS the Gaussian EMs are not simple
functions of the symplectic eigenvalue $\tilde\nu_-$ alone. Consequently,
they provide a quantification of CV entanglement of GMEMS inequivalent
to the one determined by the negativities. Furthermore, we will now show
how these result raise the problem of the ordering of two--mode Gaussian
states according to their degree of entanglement, as quantified by different
families of entanglement measures.

\section{Extremal ordering of \newline two--mode Gaussian
states}\label{secorder}

\begin{figure}[t!]
\includegraphics[width=8cm]{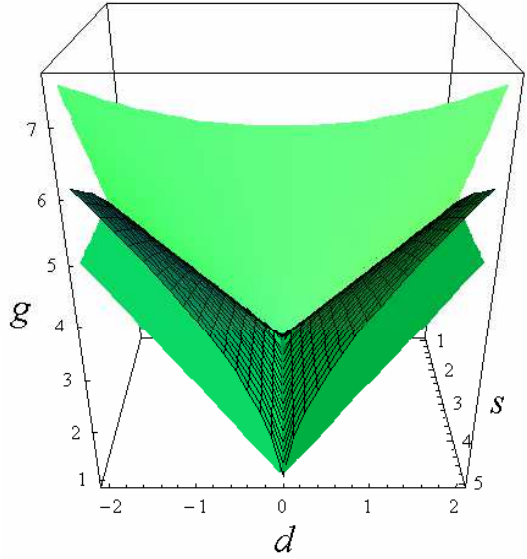}
\caption{(color online). Comparison between the ordering induced
by Gaussian EMs on the classes of states with extremal (maximal
and minimal) negativities. This {\em extremal ordering} of the set
of entangled two--mode Gaussian states is studied in the space of
the CM's parameters $\{s,d,g\}$, related to the global and local
purities by the relations $\mu_1 = (s+d)^{-1}$, $\mu_2 =
(s-d)^{-1}$ and $\mu = g^{-1}$.  The intermediate, meshed surface
is constituted by those global and local mixednesses such that the
Gaussian EMs give equal values for the corresponding GMEMS (states
of maximal negativities) and GLEMS (states of minimal
negativities). Below this surface, the extremal ordering is
inverted (GMEMS have less Gaussian EM than GLEMS). Above it, the
extremal ordering is preserved (GMEMS have more Gaussian EM than
GLEMS). However, it must be noted that this does not exclude that
the individual orderings induced by the negativities and by the
Gaussian EMs on a pair of non-extremal states may still be
inverted in this region. Above the uppermost, lighter surface,
GLEMS are separable states, so that the extremal ordering is
trivially preserved. Below the lowermost, darker surface, no
physical two-mode Gaussian states can exist. All the quantities
plotted are dimensionless.} \label{figorder3d}
\end{figure}

Entanglement is a physical quantity. It has a definite mathematical
origin within the framework of quantum mechanics, and its conceptual
meaning in the end stems from and is rooted in the existence of the
superposition principle. Further, entanglement has a fundamental
operative interpretation as that resource that in principle
enables information processing and communication in better-than-classical
realizations \cite{telepoppate}. One would then expect that, picking
two states $\varrho_A$ and $\varrho_B$ out of a certain (subset of) Hilbert
space, the question ``Is $\varrho_A$ more entangled than
$\varrho_B$?'' should have a unique, well-defined answer, independent of
the measure that one chooses to quantify entanglement. But, contrary
to the common expectations, this is generally {\em not} the case for
mixed states. Different measures of entanglement will in general induce different, inequivalent
{\em orderings} on the set of entangled states belonging to a given
Hilbert space \cite{ordering}, as they usually measure different aspects
of quantum correlations existing in generic mixed states.

In the context of CV systems, when one restricts to symmetric,
two--mode Gaussian states, which include all pure states,
the known computable measures of entanglement all correctly
induce {\em the same} ordering on the set of entangled states. We will now
show that, indeed, this nice feature is not preserved moving to
mixed, nonsymmetric two-mode Gaussian states.
We aim at comparing Gaussian EMs and
negativities on the two extremal classes of two--mode Gaussian
states \cite{extremal}, introducing thus the concept of {\em
extremal ordering}. At fixed global and local purities, the
negativity of GMEMS (which is the maximal one) is obviously always
greater than the negativity of GLEMS (which is the minimal one). If
for the same values of purities the Gaussian EMs of GMEMS are larger
than those of GLEMS, we will say that the extremal ordering is
preserved. Otherwise, the extremal ordering is inverted. In this
latter case, which is clearly the most intriguing, the states of
minimal negativities are more entangled, with respect to Gaussian
EMs, than the states of maximal negativities, and the inequivalence
of the orderings, induced by the two different families of
entanglement measures, becomes manifest.

The problem can be easily stated. By comparing $m_{opt}^{_{\rm
GLEMS}}$ from \eq{m2glems} and $m_{opt}^{_{\rm GMEMS}}$ from
\eq{m2gmems}, one has that in the range of global and local purities,
or, equivalently, of parameters $\{s,d,g\}$, such that
\begin{equation}\label{preserved}
m_{opt}^{_{\rm GMEMS}} \ge m_{opt}^{_{\rm GLEMS}}\; ,
\end{equation}
the extremal ordering is preserved. When \ineq{preserved} is violated,
the extremal ordering is inverted. The boundary between
the two regions, which can be found imposing the equality
$m_{opt}^{_{\rm GMEMS}} = m_{opt}^{_{\rm GLEMS}}$, yields the range
of global and local purities such that the corresponding GMEMS and
GLEMS, despite having different negativities, have equal Gaussian
EMs. This boundary surface can be found numerically, and the result
is shown in the 3D plot of Fig. \ref{figorder3d}.

\begin{figure}[t!]
\includegraphics[width=6.2cm]{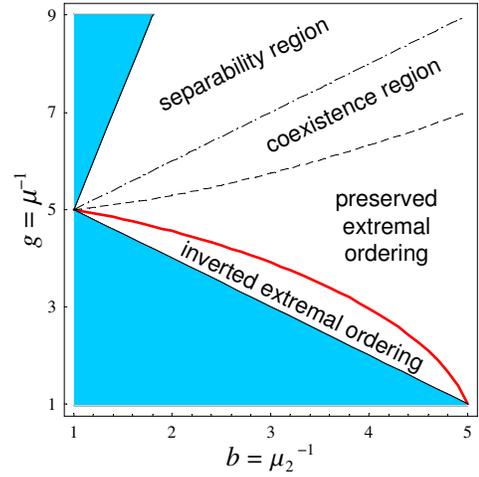}
\caption{(color online). Summary of entanglement properties of
two--mode Gaussian states, in the projected space of the local
mixedness $b = \mu_2^{-1}$ of mode $2$, and of the global mixedness
$g = \mu^{-1}$, while the local mixedness of mode $1$ is kept fixed
at a reference value $a = \mu_1^{-1} = 5$.
Below the thick curve, obtained imposing the
equality in \ineq{preserved}, the Gaussian EMs yield GLEMS more
entangled than GMEMS, at fixed purities: the extremal ordering is
thus inverted. Above the thick curve, the extremal ordering is
preserved. In the coexistence region (see Ref. \cite{prl}), GMEMS
are entangled while GLEMS are separable. The boundaries of this
region are given by \eq{glement} (dashed line) and \eq{gmement}
(dash-dotted line). In the separability region, GMEMS are separable
too, so all two--mode Gaussian states whose purities lie in that
region are not entangled. The shaded regions cannot contain any
phisical two-mode Gaussian state. All the quantities plotted are dimensionless.}
\label{figorder2d}
\end{figure}

One can see, as a crucial result, that a region where the extremal
ordering is inverted does indeed exist. The Gaussian EMs and the
negativities are thus definitely {\em not} equivalent for the
quantification of entanglement in nonsymmetric two--mode Gaussian
states. The interpretation of this result is quite puzzling. On
the one hand, one could think that the ordering induced by the
negativities is a natural one, due to the fact that such measures
of entanglement are directly inspired by the necessary and
sufficient PPT criterion for separability. Thus, one would expect
that the ordering induced by the negativities should be preserved
by any {\em bona fide} measure of entanglement, especially if one
considers that the extremal states, GLEMS and GMEMS, have a clear
physical interpretation \cite{extremal}. Therefore, as the
Gaussian EoF is an upper bound to the true EoF, one could be
tempted to take this result as an evidence that the Gaussian EoF
overestimates the true EoF, at least for GLEMS, and that,
moreover, the true EoF of GLEMS should be lower than the true EoF
of GMEMS, at fixed values of the purities. If this were the case,
the true EoF would not coincide with the Gaussian EoF, whose
evaluation would consequently necessarily involve a decomposition
over non-Gaussian states. However, this is only a
qualitative/speculative argument: proving or disproving that the
Gaussian EoF is the true EoF for any two--mode Gaussian state is
still an open question under lively debate \cite{openprob}.

On the other hand, one could take the simplest
discrete-variable instance, constituted by a two--qubit system, as a
test-case for comparison. There, although
for pure states the negativity coincides with the
concurrence, an entanglement monotone equivalent to the EoF for
all states of two qubits \cite{Wootters}, the two measures cease to be equivalent
for mixed states, and the orderings they induce on the set of
entangled states can be different \cite{frank}.
This analogy seems to support again the stand that, in the arena of mixed states,
a unique measure of entanglement is a {\it chimera} and cannot really be expected,
due to the different operative meanings and
physical processes (in the cases when it has been possible
to identify them) that are associated to each definition: one could
think, for instance, of the operative difference existing
between the definitions of distillable entanglement and
entanglement cost. In other words, from this point of view,
each inequivalent measure of entanglement introduced for mixed states
should capture physically distinct aspects of quantum correlations
existing in these states. Then, joining this kind of outlook,
one could hope that the Gaussian EMs might still be considered
as proper measures of CV entanglement, especially if one were
able to prove the conjecture that the Gaussian EoF
is the true EoF for a broader class of Gaussian states beyond the
symmetric ones. One could then live on with the existence of
inverted orderings of entangled states, and see it as a not so
annoying problem.

Whatever be the case, we have shown that
two different families of measures of CV entanglement
can induce different orderings on the set of two--mode entangled
states. This is more clearly illustrated in Fig. \ref{figorder2d},
where we keep fixed one of the local mixednesses and we classify, in
the space of the other local mixedness and of the global mixedness,
the different regions related to entanglement and extremal ordering
of two--mode Gaussian states, improving and completing a similar
diagram previously introduced in Ref. \cite{prl} to describe separability in
the space of purities.

\section{Gaussian measures of entanglement \newline versus
negativities}\label{secvs}

In this section we wish to give a more direct comparison of
the two families of entanglement measures for two--mode Gaussian
states. In particular, we are interested in finding the
maximum and minimum values of one of the two measures, if the other is
kept fixed. A very similar analysis has been performed by Verstraete {\em
et al.} \cite{frank}, in their comparative analysis of the
negativity and the concurrence for states of two-qubit systems.

Here it is useful to perform the comparison directly between the
symplectic eigenvalue $\tilde\nu_-(\sig)$ of the partially
transposed CM $\tilde \sig$ of a generic two--mode Gaussian state
with CM $\sig$, and the symplectic eigenvalue
$\tilde\nu_-(\sig^P_{opt})$ of the partially transposed CM $\tilde
\sig^{P}_{opt}$ of the optimal pure state with CM $\sig^P_{opt}$,
which minimizes \eq{Gaussian EMm}. In fact, the negativities are all
monotonically decreasing functions of $\tilde\nu_-(\sig)$, while the
Gaussian EMs are all monotonically decreasing functions of
$\tilde\nu_-(\sig^P_{opt})$.

To start with, let us recall once more that for pure states and for mixed
symmetric states (in the set of two--mode Gaussian states), the two
quantities coincide. For nonsymmetric states, one can immediately
prove the following bound
\begin{equation}\label{ubound}
\tilde\nu_-(\sig^P_{opt}) \le \tilde\nu_-(\sig)\,.
\end{equation}
In fact, from \eq{Gaussian EMm}, $\sig^P_{opt} \le \sig$
\cite{GEOF}. For positive matrices, $A \ge B$ implies $a_k \ge b_k$,
where the $a_k$s (resp. $b_k$s) denote the ordered symplectic eigenvalues of
$A$ (resp. $B$) \cite{giedkeqic}. Because the ordering $A \ge B$ is preserved
under partial transposition, \ineq{ubound} holds true. This fact
induces a characterization of symmetric states, which saturate
\ineq{ubound}, as the two--mode Gaussian states with {\em minimal}
Gaussian EMs at fixed negativities.

\begin{figure}[t!]
\includegraphics[width=8.5cm]{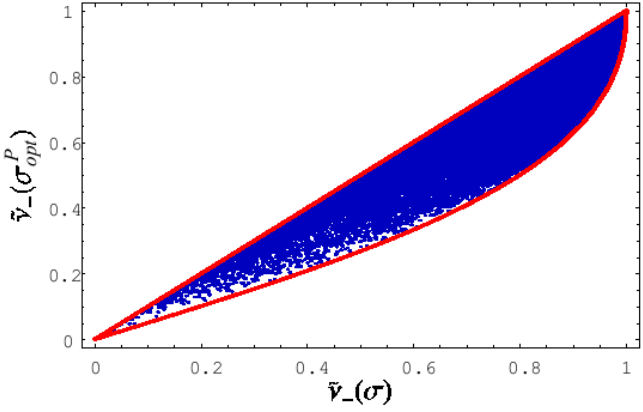}
\caption{(color online). Comparison between Gaussian EMs and
negativities for two--mode Gaussian states. On the horizontal axis
we plot the symplectic eigenvalue $\tilde\nu_-(\sig)$ of the
partially transposed CM $\tilde \sig$ of a generic two--mode
Gaussian state with CM $\sig$. On the vertical axis we plot the
symplectic eigenvalue $\tilde\nu_-(\sig^P_{opt})$ of the partially
transposed CM $\tilde \sig^{P}_{opt}$ of the optimal pure state with
CM $\sig^P_{opt}$, which minimizes \eq{Gaussian EMm}. The
negativities are all monotonically decreasing functions of
$\tilde\nu_-(\sig)$, while the Gaussian EMs are all monotonically
decreasing functions of $\tilde\nu_-(\sig^P_{opt})$. The equation of
the two boundary curves are obtained from the saturation of
\ineq{ubound} (upper bound) and \ineq{lbound} (lower bound),
respectively. The dots represent 50~000 randomly generated CMs of
two--mode Gaussian states. Of up to $1$ million random CMs, none has
been found to lie below the lower solid-line curve, enforcing the conjecture
that it be an absolute  boundary for all two--mode Gaussian states.
All the quantities plotted are dimensionless.} \label{figfrank}
\end{figure}

It is then natural to raise the question whether an upper bound
on the Gaussian EMs at fixed negativities exists as well. It seems hard to
address this question directly, as one lacks a closed expression for
the Gaussian EMs of generic states. But we can promptly give partial
answers if we restrict to the classes of GLEMS and of GMEMS, for
which the Gaussian EMs have been explicitely computed in the previous section.

Let us begin with the GLEMS. We can compute the squared
symplectic eigenvalue $\tilde\nu_-^2(\sig^{_{\rm GLEMS}}) =
 {\left[4 (s^2 +d^2) - g^2-1 -
 \sqrt{\left(4 (s^2 +d^2) - g^2-1\right)^2 - 4 g^2}\right]/2}$. Next,
 we can reparametrize the CM (obtained by \eq{cpm} with $\lambda =
 -1$) to make $\tilde\nu_-$ appear explicitely, namely
$g = \sqrt{\tilde\nu_-^2
[4(s^2+d^2)-1-\tilde\nu_-^2]/(1+\tilde\nu_-^2)}$. At this point, one
can study the piecewise function $m_{opt}^{_{\rm GLEMS}}$ from
\eq{m2glems}, and find out that it is a convex function of $d$ in
the whole space of parameters corresponding to entangled states. Hence,
$m_{opt}^{_{\rm GLEMS}}$, and thus the Gaussian EM, is maximized at
the boundary $|d| = (2 \tilde\nu_- s - \tilde\nu_-^2 - 1)/2$,
resulting from the saturation of \ineq{gmbound}. The states
maximizing Gaussian EMs at fixed negativities, if we restrict
to the class of GLEMS, have then to be found in the subclass of
GMEMMS (states of maximal negativity for fixed marginals
\cite{extremal}, defined after \ineq{gmbound}), depending on the
parameter $s$ and on the eigenvalue $\tilde\nu_-$ itself, which
completely determines the negativity). For these states,
\begin{equation}\label{mgm}
    m^{_{\rm GMEMMS}}_{opt} (s, \tilde \nu_-) =
    \left(\frac{2s}{1-\tilde\nu_-^2 + 2 \tilde\nu_- s}\right)^2\,.
\end{equation}
The further optimization over $s$ is straightforward because
$m^{_{\rm GMEMMS}}_{opt}$ is an increasing function of $s$, so its
global maximum is attained for $s \rightarrow \infty$. In this
limit, one has simply
\begin{equation}\label{mmax}
    m^{_{\rm GMEMMS}}_{\max} (\tilde \nu_-) =\frac{1}{\tilde \nu_-^2}\,.
\end{equation}
From \eq{nutopt}, one thus finds that for all GLEMS the following
bound holds
\begin{equation}\label{lbound}
\tilde\nu_-(\sig^P_{opt}) \ge
\frac{1}{\tilde\nu_-(\sig)}\Big(1-\sqrt{1-\tilde\nu_-^2(\sig)}\Big)\,.
\end{equation}

One can of course perform a similar analysis for GMEMS. But, after
analogous reasonings and computations, what one finds is exactly the
same result. This is not so surprising, keeping in mind that GMEMS,
GLEMS and all two--mode Gaussian states with generic $s$ and $d$ but
with global mixedness $g$ saturating \ineq{gmbound}, collapse into
the same family of two--mode Gaussian states, the GMEMMS, completely
determined by the local single--mode properties (they can be viewed
as a generalization of the pure two--mode states: the symmetric
GMEMMS are in fact pure). Hence, the bound of \ineq{lbound},
limiting the Gaussian EMs from above at fixed negativities, must
hold for all GMEMS as well.

At this point, it is tempting to conjecture that \ineq{lbound} holds
for all two--mode Gaussian states. Unfortunately, the lack of a closed,
simple expression for the Gaussian EM of a generic state makes the
proof of this conjecture impossible, at the present time.
However, one can show, by analytical power-series expansions
of \eq{mfunc}, truncated to the leading order in the infinitesimal
increments, that, for any
infinitesimal variation of the parameters of a generic CM around the
limiting values characterizing GMEMMS, the Gaussian EMs of the
resulting states lie always below the boundary imposed by the
corresponding GMEMMS with the same $\tilde\nu_-$. In this sense, the
GMEMMS are, at least, a {\em local} maximum for the Gaussian EM
versus negativity problem. Furthermore, extensive numerical
investigations of up to a million CMs of randomly generated two--mode Gaussian
states, provide confirmatory evidence that GMEMMS attain indeed the
{\em global} maximum (see Fig. \ref{figfrank}). We can thus quite
confidently conjecture, however, at the moment, without
a complete formal proof of the statement, that GMEMMS, in the limit of infinite
average local mixedness ($s \rightarrow \infty$), are the states of
maximal Gaussian EMs at fixed negativities, among {\em all}
two--mode Gaussian states.

\begin{figure}[t]
\includegraphics[width=8.5cm]{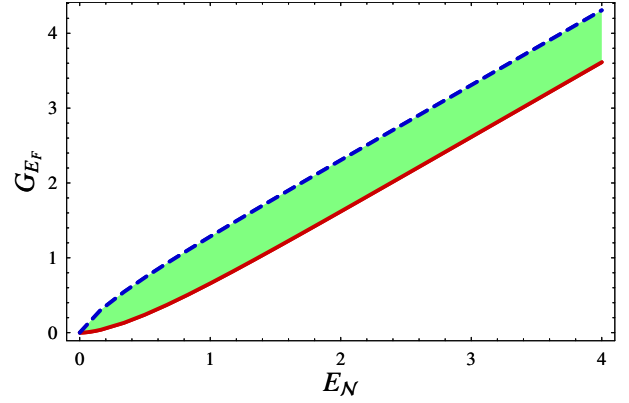}
\caption{(color online). Comparison between the Gaussian
entanglement of formation $G_{E_F}$ and the logarithmic negativity
$E_\N$ for two--mode Gaussian states. Symmetric states accomodate
on the lower boundary (solid line), determined by the
saturation of \ineq{eflower}.
GMEMMS with infinite, average local mixedness,
lie on the dashed line, whose defining equation is obtained from
the saturation of \ineq{efupper}. All GMEMS and GLEMS lie below the
dashed line. The latter is conjectured, with strong numerical
support, to be the upper boundary for the Gaussian EoF of all
two--mode Gaussian states, at fixed negativity. All the quantities
plotted are dimensionless.} \label{geofvslone}
\end{figure}

A direct comparison between the two prototypical representatives of
the two families of entanglement measures, respectively the Gaussian
EoF $G_{E_F}$ and the logarithmic negativity $E_\N$, is plotted in
Fig. \ref{geofvslone}. For any fixed value of $E_\N$, \ineq{ubound}
provides in fact a rigorous lower bound on $G_{E_F}$, namely
\begin{equation}\label{eflower}
G_{E_F} \ge h[\exp(-E_\N)]\,,
\end{equation}
while \ineq{lbound} provides the conjectured lower bound
\begin{equation}\label{efupper}
G_{E_F} \le
h\left[\exp(E_\N)\left(1-\sqrt{1-\exp(-2E_\N)}\right)\right]\,,
\end{equation}
where we exploited Eqs.~(\ref{ennu},\ref{geofm}) and $h[x]$ is given
by \eq{hentro}.

The existence of lower and upper bounds on the Gaussian EMs at fixed
negativities (the latter strictly proven only for extremal states),
limits to some extent the inequivalence arising between the two families of
entanglement measures, for nonsymmetric two--mode Gaussian states.

\section{Summary and Outlook}\label{secconcl}

In this work we focused on the simplest conceivable states of a
bipartite CV system: two--mode Gaussian states. We have shown that,
even in this simple instance, the theory of quantum entanglement
hides several subtleties and reveals some surprising aspects.
In particular, we have studied the relations existing between
different computable measures of entanglement, showing how
the negativities (including the standard logarithmic negativity) and
the Gaussian convex-roof extended measures (Gaussian EMs, including
the Gaussian entanglement of formation \cite{GEOF}) are inequivalent
entanglement quantificators for nonsymmetric two--mode Gaussian
states. We have computed Gaussian EMs explicitely for the two
classes of two-mode Gaussian states having extremal (maximal and minimal)
negativities at fixed purities \cite{extremal}. We have highlighted how,
in a certain range of values of the global and local purities,
the ordering on the set of entangled states, as induced by the Gaussian
EMs, is inverted with respect to that induced by the negativities. The
question whether a certain Gaussian state is more entangled than
another, thus, has no definite answer, not even when only extremal
states are considered, as the answer comes to depend on the measure
of entanglement one chooses. Extended comments on the possible meanings
and consequences of the existence of inequivalente orderings of
entangled states have been given in Section \ref{secorder} and in
Section \ref{secvs}. Furthermore, we have proven the existence of
a lower bound holding for the Gaussian EMs at fixed negativities,
and that this bound is saturated by two--mode symmetric
Gaussian states. Finally, we have provided some strong
numerical evidence, and partial analytical proofs
restricted to extremal states, that an upper bound on the Gaussian
EMs at fixed negativities exists as well, and is saturated
by states of maximal negativity for given marginals,
in the limit of infinite average local mixedness.

We believe that our results will raise renewed interest
in the problem of the quantification of entanglement in CV systems,
which seemed fairly well understood in the special instance of
two--mode Gaussian states. Moreover, we hope that the present work
may constitute a first step toward the solution of more general problems
concerning the entanglement of Gaussian states, such as the computation of the
entanglement of formation for generic two--mode Gaussian states
\cite{openprob}, and the proof of its identity with the
Gaussian EoF in a larger class of Gaussian states beyond the symmetric instance.
On the other hand, the explicit expressions, computed in the present work,
now available for the Gaussian EoF of GMEMS and GLEMS, might serve as well as a
basis to find an explicit counterexample to the conjecture that the
decomposition over all pure Gaussian states, in the definition of the
EoF, is the optimal one for all two--mode Gaussian states.

Finally, the results collected in the present work
might prove useful as well in
the task of quantifying multipartite entanglement of Gaussian states.
For instance, we should mention here that
any two--mode reduction of a pure three--mode Gaussian state is a GLEMS,
as a consequence of the Schmidt decomposition operated at the CM
level \cite{holewer}. Therefore, thanks to the results that we have
derived here, its Gaussian EoF can be explicitely
computed, and can be compared with the entropy of entanglement between one
reference mode and the remaining two in the global state.
One has then available the tools and can apply them
to investigate the sharing structure of multipartite CV entanglement
of three-mode, and, more generally, multimode Gaussian states \cite{contangle}.


\acknowledgements{We are grateful to O. Kr\"uger and M. M. Wolf for
fruitful discussions and for providing us with supplementary
material on the computation of the Gaussian entanglement
of formation. Financial support from CNR-Coherentia, INFN, and MIUR is acknowledged.}


\end{document}